\documentclass[twocolumn]{aastex631}
\usepackage{float}
\usepackage{xcolor}
\usepackage{soul}
\usepackage[T1]{fontenc}
\usepackage{subfigure}
\usepackage{wrapfig}
\usepackage{hyperref}
\usepackage{chngpage}

\graphicspath{}
\DeclareUnicodeCharacter{2212}{-}

\begin{document}
\title{Rapid Optical Flares in the Blazar OJ 287 on Intraday Timescales with TESS}

\author[0000-0001-8716-9412]{Shubham Kishore}
\affiliation{Aryabhatta Research Institute of Observational Sciences (ARIES), Manora Peak, Nainital 263001, India}
\affiliation{Department of Physics, DDU Gorakhpur University, Gorakhpur - 273009, India}

\author[0000-0002-9331-4388]{Alok C.\ Gupta}
\affiliation{Aryabhatta Research Institute of Observational Sciences (ARIES), Manora Peak, Nainital 263001, India}
\affiliation{Key Laboratory for Research in Galaxies and Cosmology,\\ Shanghai Astronomical Observatory, Chinese Academy of Sciences, Shanghai 200030, People's Republic of China}

\author[0000-0002-1029-3746]{Paul J.\ Wiita}
\affiliation{Department of Physics, The College of New Jersey, 2000 Pennington Rd., Ewing, NJ 08628-0718, USA}

\email{amp700151@gmail.com (SK), acgupta30@gmail.com (ACG), wiitap@tcnj.edu (PJW)}

\begin{abstract}
\noindent
 We have analyzed the optical light curves of the blazar OJ 287 obtained with the Transiting Exoplanet Survey Satellite (TESS) over about 80 days from 2021 October 13 to December 31, with an unprecedented sampling of 2 minutes. Although significant variability has been found during the entire period, we have detected two exceptional flares with flux nearly doubling and then nearly tripling over 2 days in the middle of 2021 November. We went through the light curves analysis using the excess variance, generalized Lomb-Scargle periodogram, and Continuous Auto-Regressive Moving Average (CARMA) methods, and estimated the flux halving/doubling timescales. The most probable shortest variability timescale was found to be 0.38 days in the rising phase of the first flare. We briefly discuss some emission models for the variability in radio-loud active galactic nuclei that could be capable of producing such fast flares. 
\end{abstract}

\section{Introduction}\label{sec:intro}
\noindent
The blazar subclass of radio loud (RL) active galactic nuclei (AGN)  displays flux, spectral and polarization variability throughout the electromagnetic (EM) spectrum on diverse timescales ranging from a few minutes to several years.  Blazar emission is predominantly  non-thermal. Probably  the least well understood temporal variability is observed on timescales of minutes to several hours and commonly called as microvariability \citep{1989Natur.337..627M} or intraday variability (IDV) \citep{1995ARA&A..33..163W}. Microvariability in optical flux provides a strong route toward understanding the physical processes occurring in the most compact emitting regions of blazars.\\ 
\\
\citet{1989Natur.337..627M} made the pioneering discovery of optical microvariability  in the blazar BL Lacerate, and  in the subsequent $\sim$3.5 decades, many of the brighter blazars have been observed for the study of microvariability during  thousands of observing nights using various telescopes \citep[e.g.,][and references therein]{1989Natur.337..627M,1992AJ....104...15C,1993A&A...271..344W,1996A&A...305...42H,1999A&AS..134..453S,2004MNRAS.348..176S,2002AJ....123..678Q,2006A&A...451..435M,2008AJ....135.1384G,2009ApJS..185..511P,2012MNRAS.425.3002G,2015MNRAS.450..541A,2018ApJ...863..175G,2020MNRAS.496.1430P,2021ApJS..257...41K,2021MNRAS.501.1100R,2023MNRAS.519.2796D,2023MNRAS.518.1459P,2019ApJ...877..151W,2023ApJ...951...58W}. In a statistical study of optical microvariability properties of various classes of AGN, \citet{2005A&A...440..855G} found that if a blazar is observed continuously for less than 6 h, the chance of  seeing micro-variations is $\approx$60$-$65\%, but it rises to 80$-$85\% for  nightly observations that exceed 6 h. \\

\begin{figure}[h]
    \centering
    \resizebox{3.4in}{!}{\includegraphics{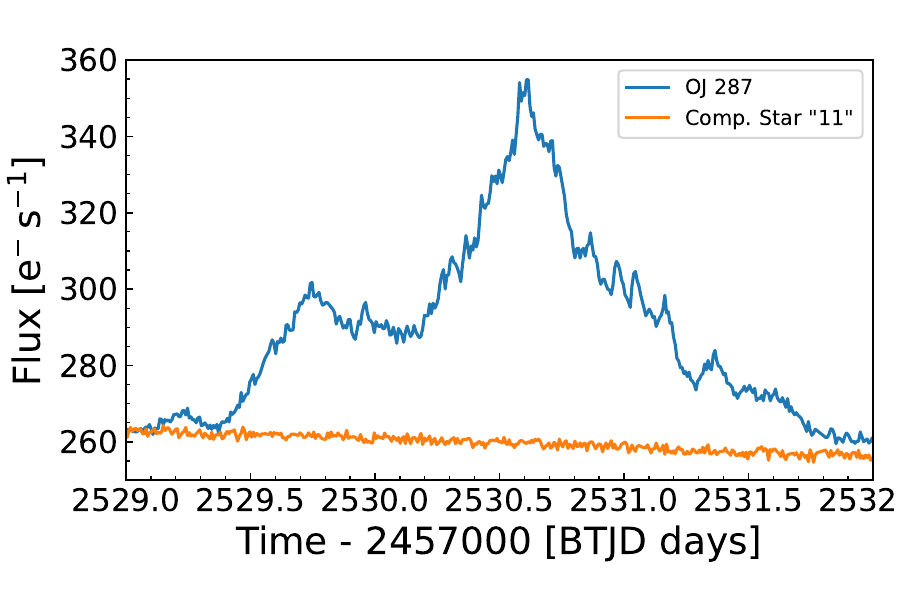}}
    \caption{ Partial Sector 45 LCs of OJ 287 and comparison star 11 during the strongest flares.}
    \label{OJ 287 comp}
\end{figure}

\begin{figure}[h]
    \centering
\resizebox{3.4in}{!}{\includegraphics{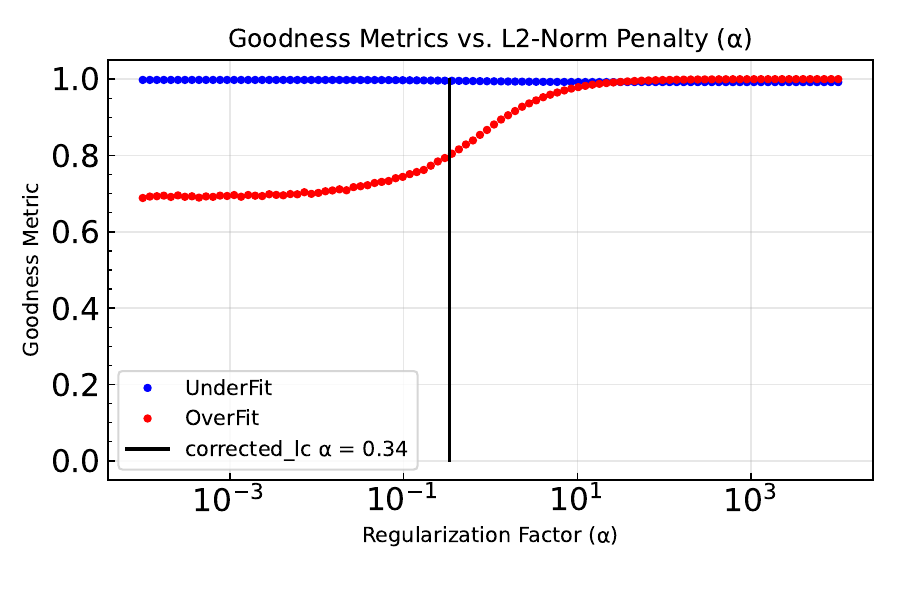}}
    \caption{Goodness metric scan plot for the TESS Sector 45 observation of OJ 287.}
    \label{GM_45}
\end{figure}

\begin{figure*}[h]
    \vspace{-1.5cm}
    \hspace{.5cm}
    \resizebox{18cm}{!}{\includegraphics[trim=0 1.cm 0 0]{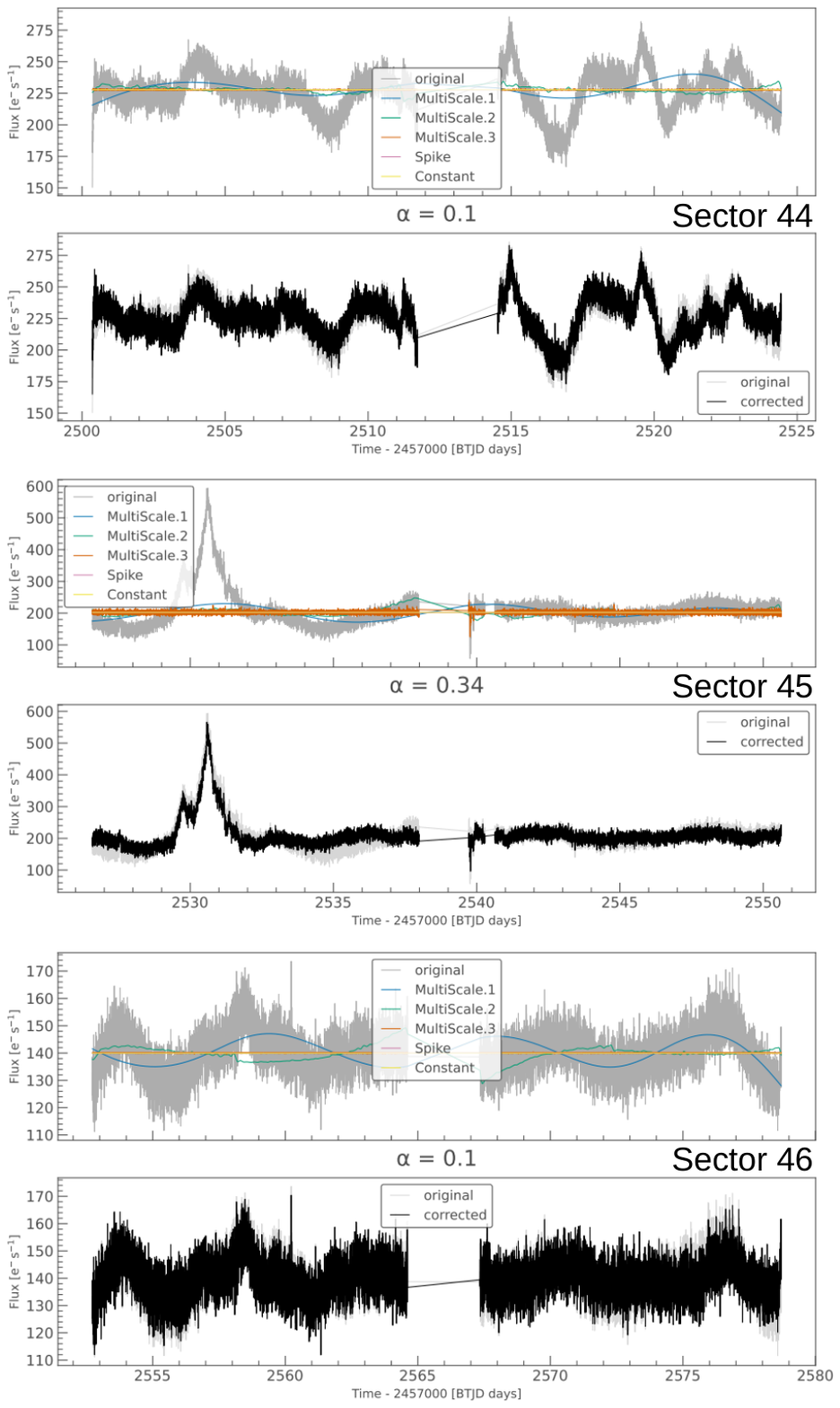}}
    \caption{Raw and reduced LCs of OJ 287 observed in all three sectors. The upper panels include the original PDCSAP fluxes and the co-trending basis vectors that are used to correct it and the bottom panels show the original and corrected LCs corresponding to each labeled sector.}
    \label{LC_all}
\end{figure*}
\begin{figure*}[t]
    \hspace{-1.cm}
    \resizebox{18.5cm}{!}{\includegraphics{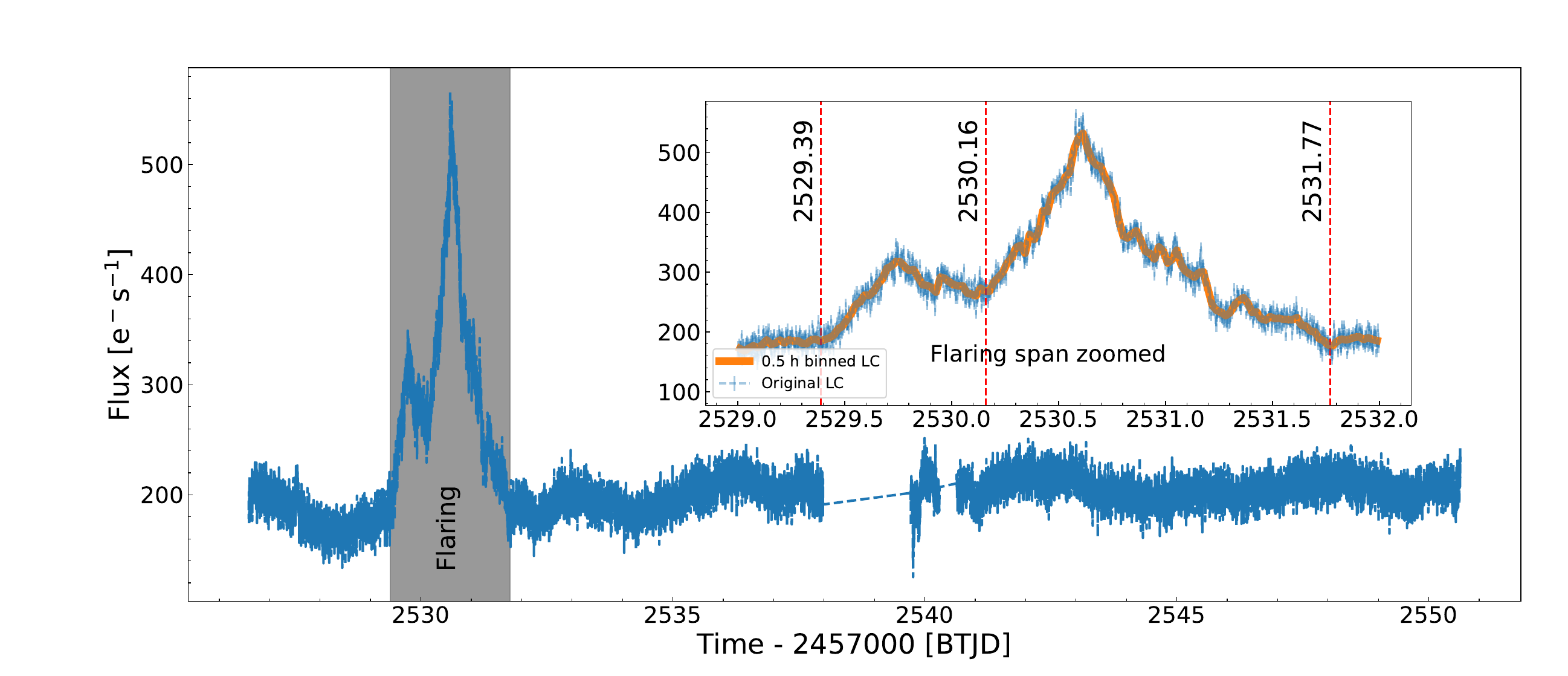}}
    \caption{The main plot includes the complete reduced Sector 45 LC. The subplot  zooms in on the flare period; for better visualization of the trend, a 0.5 h binned LC has been overplotted.}
    \label{Fig:2}\end{figure*}

\noindent
OJ 287, at redshift $z$ = 0.306 \citep{1985PASP...97.1158S}, is among the few AGN which  probably host a supermassive black hole (SMBH) binary system. OJ 287 was observed in optical bands since 1888, and by using this century-long data, \citet{1988ApJ...325..628S} discovered that the blazar shows  outbursts with a period  of $\sim$12 yrs and proposed a binary SMBH model to explain it. Because of this short orbital period, OJ 287 is thus a candidate to emit nano-hertz (nHz) gravitational waves (GWs) \citep[e.g.,][and references therein]{2021Galax..10....1V,2023MNRAS.521.6143V}. OJ 287 has been studied extensively for optical variabilities on diverse timescales, e.g., microvariability, short term variability, and long term variability \citep[e.g.,][and references therein]{1996A&A...305L..17S,1996A&A...315L..13S,2017MNRAS.465.4423G,2019AJ....157...95G,2018ApJ...863..175G,2019ApJ...877..151W,2023ApJ...951...58W}. \\
\\
 Here we consider data taken on OJ 287 over a span of about 80 days (from 2021 October 13 to 2021 December 31) by the Transiting Exoplanet Survey Satellite (TESS)\footnote{\url{http://tess.gsfc.nasa.gov}}.  Our focus is on a continuous and uniformly sampled set of observations, more precisely the first segment of sector 45 that spans $\sim$12 days with time resolution of 2 minutes. These data provide us an excellent opportunity to study optical  microvariability of a blazar with essentially uniform sampling made at the shortest time resolution and over a quite  extended duration. We found a double-peaked strong flare in the light curve (LC) that spanned about 2 days. We are unaware of  a previous clear case for such a well-resolved double-peaked flare on such timescales. OJ 287 was earlier observed from the Kepler satellite in its K2 mission phase for a  continuous period of 75 days (2015 April 27 to 2015 July 10) and this uniformly sampled optical LC  displayed several significant flares \citep{2018ApJ...863..175G,2019ApJ...877..151W}, but none as fast with as large an amplitude as those we see in these TESS data. \citet{2023ApJ...951...58W} analyzed a subsequent K2 observation of the source taken for 51 days (from 2018 May 13 to 2018 July 2) and also reported multiple rapid, but not quite as strong, flares during this epoch. \\ 
\\
In Section 2, we describe data acquisition and reduction. In Section 3, we explain the data analysis techniques we used and present results. A discussion is provided in Section 4.

\section{Data Acquisition and Reduction}
\begin{figure*}
    \resizebox{17.6cm}{!}{\includegraphics[trim=1.8cm .5cm 2cm .8cm]{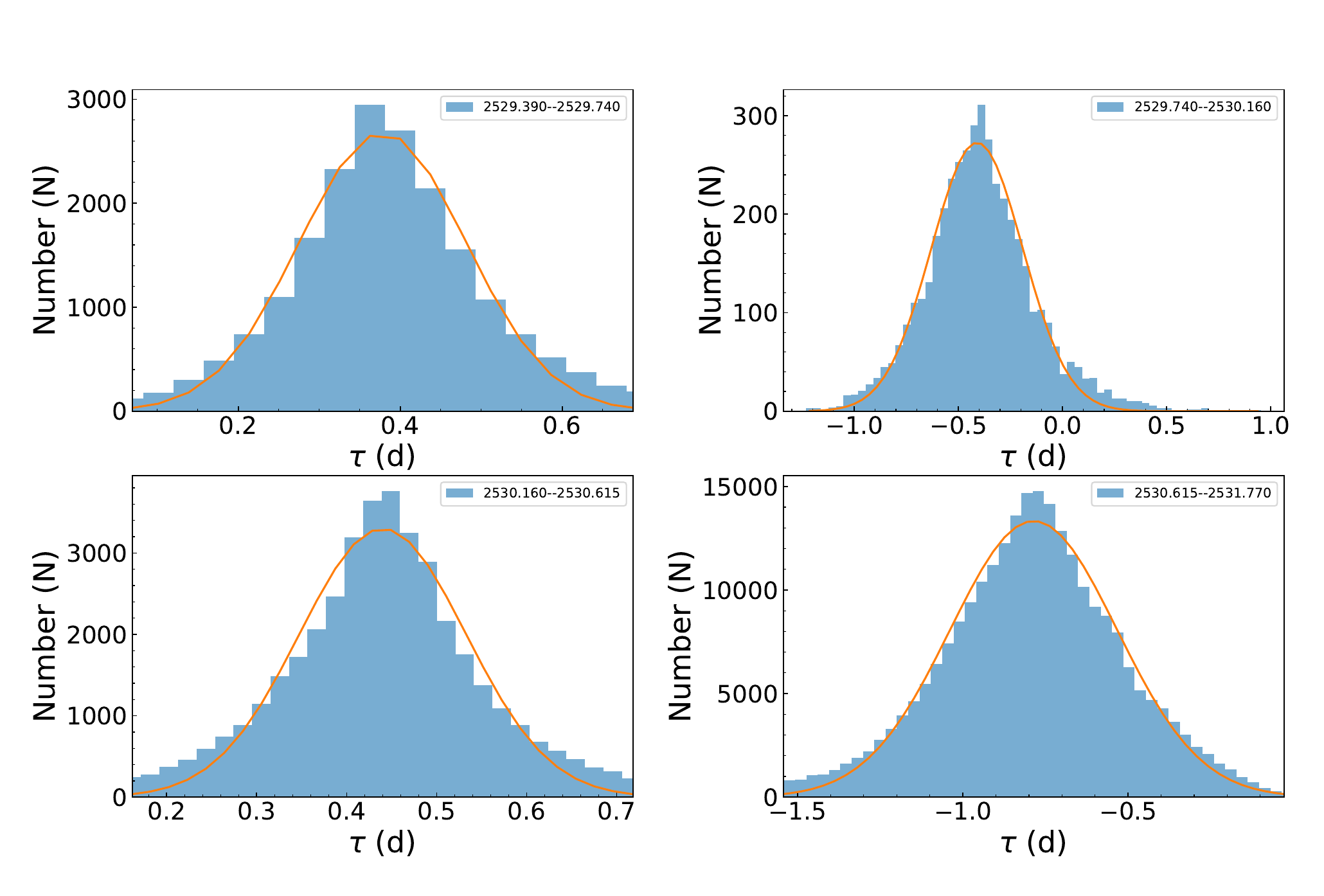}}
    \caption{Each panel shows the distribution of doubling/halving timescales corresponding to the labeled time spans. Positive values of $\tau$ indicate the rising phases and negative ones the declining phases of the flares. The orange curve in each panel shows the Gaussian fit determining the best value of $\tau$.}
    
        \label{tau_dist}
\end{figure*}
\begin{figure*}[t]
    \vspace{-.12cm}

    \resizebox{18.5cm}{!}{\includegraphics[trim=1.cm .9cm 0 0.5cm]{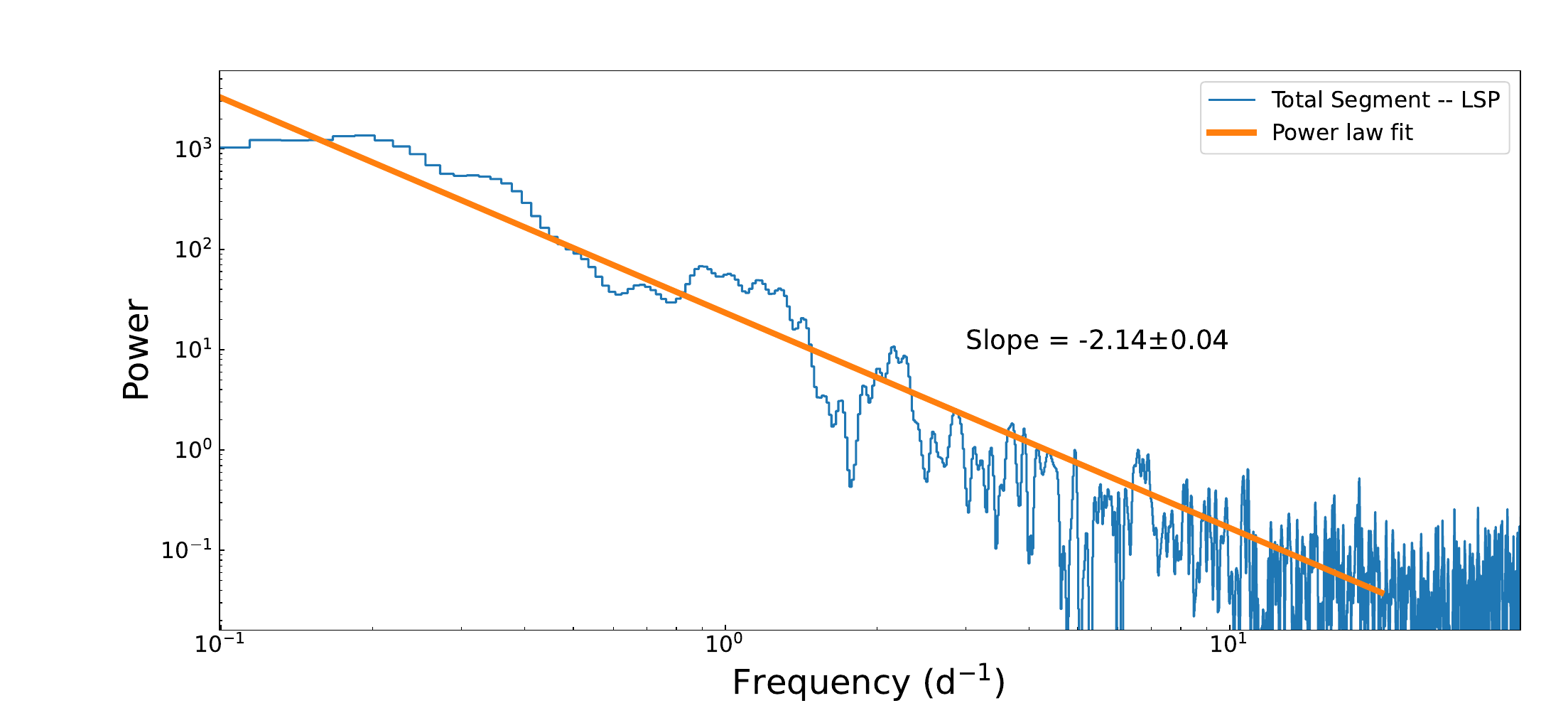}}
    \resizebox{18.5cm}{!}{\includegraphics[trim=1cm 1.cm 0 0.5cm]{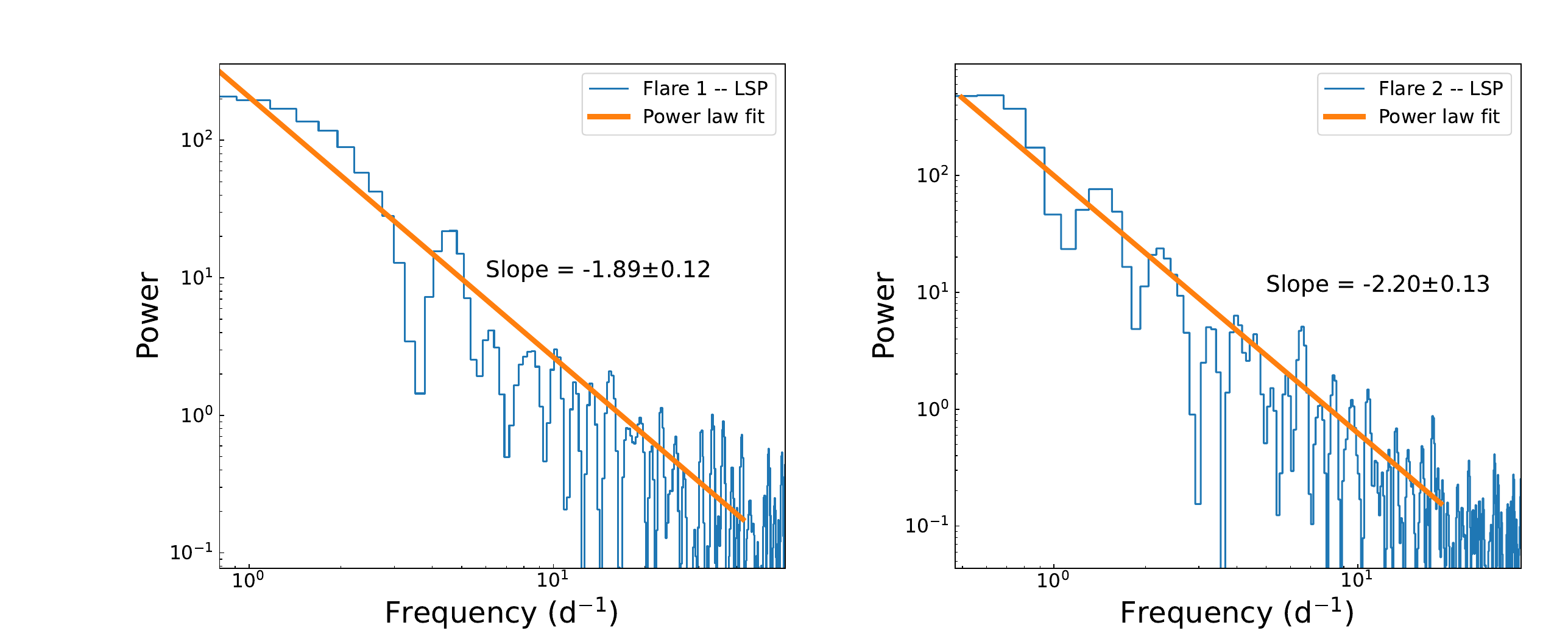}}
    \vspace{.15cm}
    \caption{The panels show the Lomb-Scargle periodograms corresponding to the labeled pieces of the first segment of the Sector 45 LC with their PSD slopes.}
    \label{psd}
\end{figure*}
\noindent

Aside from some data gaps of 1--2 days related to telemetry, OJ 287 was observed essentially continuously by TESS;  its detector bandpass spans the range 600 -- 1000 nm and is centered at the traditional Cousins I-band\footnote{\url{https://heasarc.gsfc.nasa.gov/docs/tess/the-tess-space-telescope.html}}.  These observations spanned  $\sim$80 days  across sectors 44, 45 and 46,  with a cadence of 2 minutes.  We have used the PDCSAP\_FLUX \citep{2016SPIE.9913E..3EJ} values for reduction as described in \citet{2023ApJ...943...53K} and have followed the reduction procedure given there. Hence, we refer the readers to that paper for the nomenclature used in this section and the optimum values found for the three parameters employed during data reduction: the two goodness metrics (overfitting and underfitting) and the regularization factor ($\alpha$). \\
\\
 Unfortunately there were no LC files for nearby comparision stars (which contain SAP and PDCSAP fluxes) from the TESS pipeline already available, because these values are made available only for pre-selected targets. TESS also created    full-frame images (FFIs) of the portions of sky observed at a cadence of 20 minutes for those sectors, so we used these FFIs to extract the SAP LCs of comparison stars 4, 10, 11 \citep{1985AJ.....90.1184S} along with that of OJ~287 to confirm that our source's variability is intrinsic.
Comparison stars 4 and 10 had substantially higher fluxes than OJ 287, so, in Figure\ \ref{OJ 287 comp} we used star 11, with a very comparable brightness, to illustrate the genuine variability of the source.  

It should be emphasized that Fig.\ \ref{OJ 287 comp} provides a preliminary version of the LC that provides a useful visual comparison, but a more fully reduced version  has been used in our actual analysis of the LC of OJ 287.\\
\\
Table \ref{tab:cal.details} includes the values of these fitting parameters obtained for reduction of PDCSAP\_FLUX of our object for each of the Sectors.
 As discussed in \cite{2023ApJ...943...53K}, the values of both the underfitting and overfitting goodness metrics should be kept at or above 0.8, consistent with the lowest possible $\alpha$ value. For sector 45, the overfitting metric was below 0.8 at $\alpha = 0.1$, so the optimum value of $\alpha$ was found to be 0.34 when the overfitting metric crosses the critical limit of 0.8. Fig.\ \ref{GM_45} illustrates these conditions for the sector 45 LC reduction.
Fig. \ref{LC_all} includes two plots for each of the Sectors. The upper panel in each set contains the raw LC and the different co-trending basis vectors used for the reduction, while the lower panel compares the raw LC and reduced LC. Variability is seen in all three of these LCs, but it is much stronger in Sector 45. In Fig. \ref{Fig:2} we show the zoomed in  Sector 45 reduced LC, showing the double-flare in the first segment of this sector in more detail.

\begin{table}[h]
\caption{Flux calibration details}
\hspace{-1.0cm} 
\resizebox{3.7in}{!}{
\begin{tabular}{c c c c}\hline\hline
 Sector & $\alpha$ & Overfitting metric & Underfitting metric\\\hline
 44 & 0.10 & 0.968 & 0.999\\
 45 & 0.34 & 0.801 & 0.996\\
 46 & 0.10 & 0.984 & 1.000\\
\hline
\end{tabular}}
\label{tab:cal.details}
\end{table}

\section{Data Analysis and Results}
\subsection{Excess variance and flares}
\noindent
Flux variability is one of the fundamental properties of a typical blazar across all EM bands; however, an additional variance is present in all LCs from the measurement errors in the observations.  The excess variance method incorporates the measurement errors as well in quantifying the variability. The fractional root-mean-square variability amplitude, that is the square root of normalized excess variance, and corresponding  uncertainty, have been computed as in   \citet{2003MNRAS.345.1271V}. In our analysis of $F_{var}$ for the two flares in the first segment of Sector 45, the average flux counts after removing the two flares from the segment (where the determination of the start and end of the flares was done visually,  selecting the  minima in the LC nearest to the flaring span) was used as the baseline flux. This value came out to be $\sim$193 e$^{-}$/s. The maximum flux values during the two flares are $\sim$349  e$^{-}$/s and $\sim$566 e$^{-}$/s, respectively, corresponding to nominal increases of $\sim$81\% and $\sim$194\%, respectively. 
With this baseline, the LC shows an overall $F_{var}$ of 
72\% during the total flaring period. Table \ref{tab:Results} includes the $F_{var}$ obtained for the two flares along with the other parameters describing the flares.
\subsection{Variability timescales}
\noindent
We have used the  halving/doubling timescale given as 
\begin{equation}
    F(t)=F(t_0)\*2^{(t-t_0)/\tau}\ \ \ \ (t>t_0) ,
\end{equation}
where $\tau$ and $F(t)$ are the characteristic halving/doubling timescale and the flux value at time $t$. 
In our analysis, we first divided each flare into two parts: a `rising phase' and a `declining phase' and then examined each and every possible distinct pair of data points within each flare. We then imposed a selection criterion  so that only those data point pairs, where the differences in flux were greater than $3\sigma$, were considered for the estimation of $\tau$ \citep{2011A&A...530A..77F}. The doubling/halving time was estimated for each obtained pair. Thus, for each of the two phases of the two flares, the obtained ensemble of distinct pairs led to a distribution of the timescale.
Fig.\ \ref{tau_dist} shows the distribution of the $\tau$ estimates obtained for the two pairs of rising and declining phases and Table \ref{tab:Results} includes the most probable timescales  with the corresponding uncertainties (after fitting the distribution with a Gaussian function) for each of the four flaring phases, as well as the times of the two peaks.
 It should be noted that there is a tiny positive fluctuation during the declining phase of the LC during the first flare between epochs 2529.921 to 2529.953; so we dropped this small rising fluctuation from the declining phase while evaluating $\tau$. The two portions of the declining phase, from 2529.740 to 
 2529.921 and from 2529.953 to 2530.160, led to separate sets of $\tau$ values. These two sets of  $\tau$ values were combined to give the composite timescale distribution of this declining phase, shown in the upper right panel of Fig.\ \ref{tau_dist}. 
 \\

\begin{table*}[]
\vspace{.35cm}
\caption{Flare characteristics} 
\begin{adjustwidth}{-.9in}{}
\resizebox{8in}{!}{
    \begin{tabular}{c c c c c c c c c}\hline \hline
         Flare & Flare peak & Flare peak& Difference of flare peak  & $F_{var}$& $\tau$ (days) & $\tau$ (days)& Spectral index & KS$-$test    \\
               & epoch [BTJD]& flux (e$^-$/s) & and baseline  (e$^-$/s)  & (\%) &  (rising phase) &  (declining phase) &($\alpha$) & $p_{value}$ \\
    \hline
          1 & 2529.74 & 349 & 156 & 42.9 & 0.38 $\pm$0.10  & $-$0.41$\pm$0.21  & $-1.89 \pm 0.12$ & 0.91\\
          2 & 2530.58 & 566 & 373 & 82.4 & 0.44 $\pm$0.09 & $-$0.79 $\pm$0.25 & $-2.20 \pm 0.13$ & 0.98 \\
    \hline    
    \end{tabular}}
\end{adjustwidth}
    \label{tab:Results}
\end{table*}

\begin{table}[]
\vspace{.35cm}
\caption{Segment-wise PSD  and CARMA results}

\begin{adjustwidth}{-.55in}{}
\resizebox{3.95in}{!}{
\begin{tabular}[c]{c c c c c}
\hline \hline
  Sector/    & Cut-off  & Spectral index & KS$-$test& CARMA\\ 
   Segment     & freqs. ($d^{-1}$) &($\alpha$)& $p_{value}$&($p,q$)\\\hline
   44/1 & 0.25$-$7.00 & $-$1.99 $\pm$ 0.09 & 0.93 &(1,0)\\ 
   44/2 & 0.15$-$7.00 & $-$2.56 $\pm$ 0.09 & 0.99 &(1,0)\\ 
   45/1 & 0.07$-$20.0 & $-$2.14 $\pm$ 0.04 & 0.95 &(1,0)\\ 
   45/2 & 0.08$-$10.0 & $-$1.41 $\pm$ 0.06 & 1.00 &(1,0)\\ 
   46/1 & 0.04$-$7.00 & $-$1.56 $\pm$ 0.07 & 0.95 &(2,1)\\ 
   46/2 & 0.08$-$10.0 & $-$1.05 $\pm$ 0.06 & 1.00 &(2,0)\\ 
 \hline 
 \end{tabular}}
 \end{adjustwidth}
    \label{tab:Spectral index}
\end{table}

\subsection{Periodograms}
\noindent
Apart from the variability timescales that may be associated with the size of the emitting region, the spectral index of the periodogram, or the power spectral density (PSD) slope, can yield information about the source of the variability. Various physical models naturally yield somewhat different ranges of spectral indices \citep[e.g.,][]{2016ApJ...820...12P,2019ApJ...877..151W}.
We have used the generalized Lomb-Scargle periodogram as in
\citet{2023ApJ...943...53K}
for the analysis of all six segments of the three sectors, as well as the 
first flare and the second flare of Sector 45 individually. As usual, the PSDs so obtained flatten to instrumental 
white noise in the high frequency regime, so we used  cut-off  frequencies \citep{2008Natur.455..369G,2009A&A...506L..17L}  for each segment, which varied somewhat,
 while conservatively fitting the PSDs in the red noise  region with power laws ($P(\nu)=A\nu^\alpha$). 
  To test the power-law fit to the PSD, we followed the approach of \citet{2005A&A...431..391V}.
 This involved considering twice the ratio of the periodogram to the fitting model at each frequency. The set of these ratios were used to form a cumulative distribution function (CDF). Ideally this CDF should follow that of a $\chi_2^2$ ($\chi^2$ distribution with two degrees of freedom) \citep{2005A&A...431..391V}. Under the null hypothesis that these two distributions are the same, Kolmogorov-Smirnov (KS) tests were performed for each PSD fitting. The p-values (probability of not discarding the null hypothesis) were evaluated for each comparison of segment-wise fits, and the  high p-values we obtained (given in Table \ref{tab:Spectral index})  do indicate good PSD fits.
  The scipy{\footnote{\url{https://scipy.org/}}} python package was used for the goodness of fit estimation,
  following the steps described in \citet{2005A&A...431..391V}. An oversampling by a factor of 5 has been employed in the Lomb-Scargle periodogram calculation as there is a paucity of data-points in the low frequency red-noise region.\\
 \\
 Fig.\ \ref{psd} 
 displays the PSD behaviors of  the first segment of the Sector 45 LC  and Table \ref{tab:Spectral index} 
 gives the  spectral indices we obtained for all six of the segments.
The PSD slopes have  typically high values (close to or greater than 2) up through the segment including the flares, but following that, there seems to be a considerable flattening of the PSD slopes in the later segments. 
The overall PSD slopes during the flares and during the entire segment of Sector 45 agree within the errors. \\ 
\\
The PSD slope values listed in Table 3 for the several segments  cover quite a wide range. Hence they unfortunately cannot be used to eliminate any of the rather small number of models for blazar variablity that have evaluated the resulting PSDs \citep[e.g.][]{2016ApJ...820...12P,2019ApJ...877..151W,2021ApJ...912..109K}, each of which is, or appears to be, capable of yielding a span of slopes within this range. 

\subsection{CARMA Modeling}
\noindent
 Although the great majority of the analyses of AGN LCs  in the literature have focused on periodogram slopes, a more sophisticated approach to analyzing the structure of LCs involves considering autoregressive (AR), moving average (MA),   models, which in their continuous version are given the acronym, CARMA \citep[e.g.][]{2009ApJ...698..895K,2014ApJ...788...33K,2017MNRAS.470.3027K,2018ApJ...863..175G}. CARMA models employ the plausible assumption that a LC is a realization of a Gaussian noise process.  Specifically, a CARMA($p,q$) model connects the LC and its first $p$ time derivatives to the noise and its first $q$ time  derivatives (for the definitions used here, see \citet{2014ApJ...788...33K}, Equation (1) and accompanying text).  A CARMA(1,0) model is equivalent to a damped random walk, or Ornstein-Ulenbeck process, and seems to better describe the long term LCs of many quasars than does a single periodogram slope \citep[e.g.][]{2009ApJ...698..895K,2014ApJ...788...33K}.  \\
\\
 A physical interpretation of this  approach is that the AR part of the model describes the short-term memory in the system, while the MA part indicates how the amplitudes of random perturbations behave on different timescales. Hence, both the correlation structure and degree of smoothness of noisy processes can be described by CARMA models \citep[e.g.][]{2019PASP..131f3001M}. 
 \citet{2019ApJ...885...12R} found that blazar $\gamma$-ray LCs were usually better fit by the modestly more complex CARMA(2,1) models than by CARMA(1,0) ones, though the number of objects for which this analysis could be done was modest.  
 In a paper describing a very wide range of approaches to analysing Fermi-LAT $\gamma$-ray LCs of 11 blazars \citep{2020ApJS..250....1T} 
perform CARMA modeling of LCs binned into 7-day, 10-day and 14-day intervals.  Their results are basically consistent with those of \citet{2019ApJ...885...12R}; though both CARMA(1,0) and (2,1) models were most frequently optimal, occasionally (3,0) or (3,1) models were preferred.\\
\\
 We have performed CARMA analyses of these TESS LCs following the approach of \citet{Yu2022}\footnote{\url{https://github.com/ywx649999311/EzTao.git}}
 which provides a fast way to produce CARMA models.  All $(p,q)$ pairs with $1 \leq p \leq 5$ and $q < p$ were considered for each of the 6 segments of the 3 Sectors.  After randomly generating the initial parameter values for each of the CARMA ($p,q$) models, the LCs were fitted by those models.  The goodness of each fit was estimated by finding  the log-likelihood value with respect to the initial LC, and the best fit was taken to be the one that maximized that quantity.  The resulting best fitting ($p,q$) values are given in the last column of Table \ref{tab:Spectral index}.  We see that both Sectors 44 and 45 can be characterized by a CARMA(1,0), or a damped random walk model, though the last Sector 46 prefers models with $p=2$.

\section{Discussion}
\noindent
By comparing all the TESS observations of OJ 287 from Fig.\ \ref{LC_all}, it is clear that the double peaked flare in sector 45 illustrates a strong outburst not otherwise seen during these three Sectors. From the subplot in Fig.\ \ref{Fig:2}, it is evident that the blazar flux rises during the flares almost monotonically, but the decays involve some jerks. We also notice that the rises are faster than the decays during each of the two flare phases.  \citet{1999MNRAS.306..551C} showed that symmetric light curves, with similar rise and decay timescales \citep{2010ApJ...722..520A}, are expected when the cooling time of electrons t$_{cooling}$ is significantly shorter than the light crossing time, $R/c$, with $R$ the size of the emission region. In this analysis it was assumed that the relativistic electrons are accelerated so that their energies obtain  a power-law distribution.  Asymmetric profiles, with decay times longer than rise times, result when t$_{cooling} > R/c$ because then the time scale for decline is longer than the time scale for rise of the flare \citep{1999MNRAS.306..551C,2018MNRAS.478..172L}. \\ 
\\
 OJ 287 was earlier observed by Kepler for a continuous period of 75 days (2015 April 27 to 2015 July 10). A uniformly sampled optical LC of OJ 287 from the first K2 observation detected several rapid flares, though none as significant on these short time scales as seen in Sector 45 \citep{2018ApJ...863..175G,2019ApJ...877..151W}. \citet{2018ApJ...863..175G} found that the PSD (power spectral density) of the total LC at that time was well fitted by a CARMA (4,1) model\footnote{\url{https://github.com/brandonckelly/carma_pack}}, so of a higher order than we find during the later TESS observations.  It may be worth noting, though,  that a CARMA(4,2) or (4,3) model was among the two runners-up to the best fits we found for 5 of the 6 segments.  \citet{2023ApJ...951...58W} found the PSD slope during that period to be $-2.28\pm0.17$ for long-cadence (30 minute bins) data, whereas the PSD slope was found to be somewhat steeper, $-2.65\pm0.05$, for the short-cadence (1 minute bins) data. In the later K2 observation of OJ 287 that lasted 51 days (2018 May 13 to 2018 Jul 2), the flux variability shows a similar jagged behavior as during its 2015 observation. The PSD slopes for long-cadence (30 minutes bin) data and  short-cadence (1 minute bin) data were found to be $-1.96\pm0.20$ and $-2.26\pm0.06$, respectively \citep{2023ApJ...951...58W}, and so in the same range as the first three segments of the TESS observations discussed here. \\
\\
Although the variable flux from blazars is  dominated by jet emission, accretion disks can contribute in the low-flux state of flat spectrum radio galaxies. Since OJ 287 is a BL Lac and   during this time it was in an overall intermediate brightness state (Gupta et al., in preparation) we can rule out the strong flux variation as arising from the accretion disk. Much of the optical (and other) variability in blazars can be explained by turbulence behind the shocks in a relativistic jet \citep[e.g.,][]{2014ApJ...780...87M,2016ApJ...820...12P} or in turbulence produced by magnetic reconnection \citep[e.g.,][]{2021ApJ...912..109K,2021ApJ...919..111G}. These magnetic reconnection structures in jets can lead to  very fast emission changes that typically produce flares with somewhat longer decay than rise times \citep{2021ApJ...912..109K}, as we saw in OJ 287. The very rapid and substantial flares on sub-day timescales seen here could also arise from extremely compact regions with very high Doppler factors,  as occurring in the mini-jet or jet-in-jet scenarios \citep[e.g.,][]{2008MNRAS.386L..28G,2009MNRAS.395L..29G}.  In such models, portions of the plasma in the relativistic jets are accelerated to Lorentz factors $\sim100$ through magnetocentrifugal or magnetic reconnection processes and the resulting extreme Doppler boosting can yield fast and strong flux changes. \\
\\
 We  obtained the shortest variability timescale ($\tau_{\rm min} =$ 0.38 days) in the rising phase of the first flare. To estimate an upper limit for the size of the
emission region, $R$, we apply the simple causality constraint,
\begin{equation}
R \leq \frac{c ~\tau_\mathrm{min} ~\delta}{1+z} ~,
\end{equation}
where $\delta$ represents the Doppler factor. \citet{2018ApJ...862....1C} compiled  values of $\delta$ for OJ 287 from the literature: $\delta =$ 18.9$\pm$6 and 17.0 were derived from 43 GHz radio flares of OJ 287 in 1998-2000 and 2003, respectively \citep{2005AJ....130.1418J,2009A&A...494..527H}. A more recent value of $\delta =$ 8.7 was derived from millimeterwave flares after 2007 \citep{2017ApJ...846...98J,2017MNRAS.466.4625L}. By using the complete range of Doppler factor
values, 8.7 to 18.9$\pm$6, taking $z =$ 0.306 \citep{1985PASP...97.1158S}, applying the shortest variability timescale ($\tau_{\rm min} =$ 0.38 days), and using Equation (2), we estimate the size of emission region to be in the range of $2.2 \times 10^{15}$ cm -- $6.3 \times 10^{15}$ cm.\\
\\
 We have reduced and analyzed the TESS LCs for OJ~287 spanning three consecutive Sectors (44--46), which correspond to October 13 through December 31, 2021.  Each Sector had a $\sim$2--3 day gap near its middle, so we analyzed the resulting 6 segments separately. All of them showed significant variability which can be approximately characterized as  having PSD slopes in the range $\sim -1.5$ to $\sim -2.5$ and usually being well fit by CARMA(1,0) models. Such light curves are typical of AGN, but an unexpected result was the observation of two consecutive strong flares (with flux increases of $\sim$81\% and $\sim$194\%) seen during the first segment of Sector 45.  Both of them had most probable doubling rise times around 0.4 d and decay times that were nearly as fast, indicating that this optical emission arises from a very compact region in the relativistic jet.

\section*{ACKNOWLEDGMENTS}
\noindent
 We thank the anonymous reviewers for useful comments which helped us to significantly improve the manuscript. This paper includes data collected with the TESS mission, obtained from the MAST data archive at the Space Telescope Science Institute (STScI). Funding for the TESS mission is provided by the NASA Explorer Program. STScI is operated by the Association of Universities for Research in Astronomy, Inc., under NASA contract NAS 5–26555.
ACG is partially supported by Chinese Academy of Sciences (CAS) President's International Fellowship Initiative (PIFI) (grant no. 2016VMB073). \\
\\
{\it Facility:}  Transiting Exoplanet Survey Satellite (TESS) -- The dataset used in this paper can be found in MAST: \href{https://mast.stsci.edu/portal/Mashup/Clients/Mast/Portal.html?searchQuery=%7B%22service%22:%22DOIOBS%22,%22inputText%22:%2210.17909/b3et-af14%22%7D}{10.17909/b3et-af14.} \\
\\
{\it Software:} lightkurve \citep{2018ascl.soft12013L}, 
SciPy \citep{2020SciPy-NMeth}, EzTao \citep{Yu2022}.

\bibliography{ref} 
\bibliographystyle{aasjournal}
\end{document}